\documentclass[fleqn,12pt,showkeys]{revtex4}
\usepackage{hyperref}
\usepackage{amsmath}
\begin{document}
\title{Reply to Burridge \& Linden: Hot water may freeze sooner than cold}

\author{Jonathan I. Katz}
\affiliation{Dept. of Physics and McDonnell Center for the Space Sciences\\
Washington University, St. Louis, Mo. 63130 USA}

\email{katz@wuphys.wustl.edu}


\begin{abstract}
In a recent paper in Scientific Reports, Burridge \& Linden misinterpret
the Mpemba effect as a statement about the rate of cooling of liquid
water, when it is in fact a statement about the rate of freezing of water.
Debunking an obviously absurd claim about cooling, they miss the
significant effect, its only quantitative experimental study and a
theoretical argument that explains the effect and predicts that it occurs
only for ``hard'' water (water with significant dissolved Mg and Ca
bicarbonates).  This prediction remains to be tested.
\end{abstract}
\flushbottom
\maketitle
\thispagestyle{empty}
\section*{Introduction}
Burridge \& Linden \cite{BL16} review experiments and perform a new
experiment on the cooling of hot and cold water.  Unsurprisingly, they find,
both in their new careful experiments and in the extensive but very
heterogeneous earlier literature, that in the same thermal environment
(cold air) it takes longer for hot water to cool to $0^{\,\circ}$C than
for cold water.  However, the ``Mpemba effect'' is usually taken to be
the observation, reported since ancient times and a matter of folklore, but
hardly studied quantitatively, that initially hot water \emph{freezes}
sooner than initially cold water when placed in the same environment.  This
is a quite different assertion.
\section*{Cooling Water}
It is evident from elementary thermodynamics that the version of the
``Mpemba effect'' that Burridge \& Linden are at pains to refute must be
wrong.  As long as water is in thermodynamic equilibrium (if it were not,
a temperature could not be meaningfully defined; see, however, \cite{LR16}
who suggest that the Mpemba effect is intrinsically a nonequilibrium
phenomenon, requiring macroscopic relaxation times) and all the water in a
container is at the same temperature (usually a good approximation because
of natural buoyancy-driven convection), initially warm water must pass
through all lower temperatures on its way to $0^{\,\circ}$C.   Heat flows
from the warmer water to a colder external temperature $T_{ext} <
0^{\,\circ}$C and specific heats are positive (required for thermodynamic
stability).  If the heat flow is determined by the instantaneous temperature
difference (requiring that fluid flows in the water and the environment
relax rapidly to the instantaneous temperature difference, retaining no
memory of earlier conditions), then the time to cool from an initial
temperature $T_i > 0^{\,\circ}$C to $0^{\,\circ}$C is
\begin{equation}
t_i = \int_{T_i}^{0^{\,\circ}\mathrm{C}} {F(T_{ext}-T) \over C_P}\,dT,
\end{equation}
where $F(T_{ext}-T) < 0$ is the rate of heat flow out of the water and
$C_P > 0$ (a weak function of temperature) is the heat capacity of the
water and its container.  $dT$ is negative, integrating from a higher $T_i$
to $0^{\,\circ}$C, and the integral is positive.  Then the time $t_w$ to
cool to $0^{\,\circ}$C from a warm temperature $T_w$ is related to the
time $t_c$ required to cool to $0^{\,\circ}$C from an initially cooler
temperature $T_c$, where $0^{\,\circ}\mathrm{C} < T_c < T_w$:
\begin{equation}
\begin{split}
t_w &= \int_{T_w}^{0^{\,\circ}\mathrm{C}} {F(T_{ext}-T) \over C_P}\,dT \\
&= \int_{T_w}^{T_c} {F(T_{ext}-T) \over C_P}\,dT
+ \int_{T_c}^{0^{\,\circ}\mathrm{C}} {F(T_{ext}-T) \over C_P}\,dT \\
&= \int_{T_w}^{T_c} {F(T_{ext}-T) \over C_P}\,dT + t_c \\ &> t_c.\\
\end{split}
\end{equation}
It is hardly surprising that experimental results satisfy this inequality.
\section*{The Mpemba effect}
The Mpemba effect \cite{A95,J06} is usually considered to refer to the
\emph{freezing} of water.  Because of the large latent heat of freezing,
removal of latent heat accounts for more of the time required to freeze than
removal of internal energy in the liquid phase.  In most experiments this is
the case even if the ``hot'' water is initially at temperatures close to
$100^{\,\circ}$C because at high temperatures in unsaturated atmospheres
evaporative cooling is rapid.  Most of the cooling time is spent at
temperatures comparatively close to $0^{\,\circ}$C where evaporative cooling
is unimportant and the remaining internal energy that must be removed before
the onset of freezing is small compared to the latent heat of freezing. 

There appears to be only one quantitative experimental study of the Mpemba
effect in freezing, that of Wojciechowski, Owczarek and Bednarz
\cite{WOB88}, not cited by Burridge and Linden \cite{BL16}.  These authors
found that initially warmer water froze before cooler.
\section*{Discussion}
A theoretical explanation was offered by Katz \cite{K09}.  The explanation
involves both concentration of solutes by zone refining and freezing point
depression; the reader is referred to the original paper for details.
Contrary to an assertion in Burridge \& Linden \cite{BL16}, this paper
successfully explained the effect, at least qualitatively.  It also predicts
that an Mpemba effect will be observed in ``hard'' water containing
bicarbonates of Mg and Ca.  Heating the water removes CO$_2$, turning
bicarbonates into carbonates that precipitate \cite{P55} and ``softening''
hard water.  This gives specific form to earlier suggestions that heating
somehow changes the nature of the water, so that water that has once been
heated differs from water that has not been heated, or not heated recently
(water exposed to air recaptures CO$_2$ from the atmosphere).  Burridge \&
Linden \cite{BL16} boiled all their water samples before cooling them, so it
is predicted that they would not have observed hot water to freeze before
cold water, even if they had searched for this.

Unfortunately, authors of papers of the Mpemba effect have not described the
composition or hardness of the water they used.  Many natural waters are
hard because they are obtained from aquifers in carbonate rock.  It remains
for further experiments to compare rates of freezing of warm and cold water,
soft and hard.
\section*{Author contributions}
This paper is entirely the work of the sole author, J.I.K.
\section*{Additional information}
{\bf Competing financial interests:} The author declares no competing
financial interests.


\begin{thebibliography}{9}
\bibitem{BL16} Burridge, H. C. and Linden, P. F. Questioning the Mpemba
effect: hot water does not cool more quickly than cold.  {\it Scientific
Reports\/} {\bf 6}, 37665; doi: 10.1038/srep37665 (2016).
\bibitem{LR16} Liu, A. and Raz, O. Anomalous cooling and heating --- the
Mpemba effect and its inverse. {\it Proc. Natl. Acad. Sci.\/} submitted
arXiv:1609.05271 (2016).
\bibitem{A95} Auerbach, D. Supercooling and the Mpemba effect: When hot
water freezes quicker than cold. {\it American Journal of Physics\/}, {\bf
63}, 882--885 (1995).
\bibitem{J06} Jeng, M. The Mpemba effect: When can hot water freeze faster
than cold? {\it American Journal of Physics}, {\bf 74}, 514--522 (2006).
\bibitem{WOB88} Wojciechowski, B., Owczarek, I. \& Bednarz, G. Freezing of
aqueous solutions containing gases. {\it Crystal Research and Technology\/},
{\bf 23}, 843--848 (1988).
\bibitem{K09} Katz, J. I. When hot water freezes before cold.  {\it
American Journal of Physics\/}, {\bf 77}, 27--29 (2009).
\bibitem{P55} Pauling, L. {\it College Chemistry\/} 2nd ed., W. H. Freeman
(1955).
\end{thebibliography}
\end{document}